# Shannon-inspired Statistical Computing to Enable Spintronics

Ameya D. Patil[1], Sasikanth Manipatruni[2], Dmitri E. Nikonov[2], Ian A. Young[2], Naresh R. Shanbhag[1].

[1] Department of Electrical and Computer Engineering, University of Illinois at Urbana-Champaign, Urbana, IL 61801 USA.

[2] Components Research, Intel, Hillsboro, OR, 97124, United States.

**Summary:**

Modern computing systems based on the von Neumann architecture are built from silicon complementary metal oxide semiconductor (CMOS) transistors that need to operate under practically error free conditions with 1 error in $10^{15}$ switching events. The physical dimensions of CMOS transistors have scaled down over the past five decades leading to exponential increases in functional density and energy consumption. Today, the energy and delay reductions from scaling have stagnated, motivating the search for a CMOS replacement. Of these, spintronics offers a path for enhancing the functional density and scaling the energy down to fundamental thermodynamic limits of $100kT$ to $1000kT$. However, spintronic devices exhibit high error rates of 1 in 10 or more when operating at these limits, rendering them incompatible with deterministic nature of the von Neumann architecture. We show that a Shannon-inspired statistical computing framework can be leveraged to design a computer made from such stochastic spintronic logic gates to provide a computational accuracy close to that of a deterministic computer. This extraordinary result allowing a $10^{13}$ fold relaxation in acceptable error rates is obtained by engineering the error distribution coupled with statistical error compensation.

Computing's origins in Turing's abstract universal machine[1] and its realization through von Neumann's stored program architecture[2] have required electronic devices (transistors) that switch nearly error free with less than 1 error in $10^{15}$ switching events. The geometric dimensions (feature sizes) of these devices have been scaling relentless over the past five decades leading to improved density of transistors ($10^9/cm^2$), and a reduction in energy (10000kT/event) and delay (<10 ps), while preserving their nearly error free switching behavior. This reduction in energy is due to a commensurate scaling down of the supply voltage, which has accompanied the scaling of transistor feature size. This form of scaling, referred to popularly as Moore's Law[3], has required an aggressive adoption of a number of semiconductor process technology innovations such as superior electrostatic control[4], new gate oxide materials[5] and the manipulation of the carrier transport via the use of strained semiconductors[6]. However, in spite of successful scaling of the transistor geometric dimensions, the supply voltage and frequency of operation have stagnated due to the fundamental limits to current modulation (60 mV/decade of change in current magnitude) imposed by Boltzmann distribution of electrons at room temperature. Hence it is of great interest to explore new computational devices and new models of computation that leverage the unique properties of such devices to enable continued computational scaling. In other words, the exploration of new computational devices needs to be conducted in concert with an exploration of new models of computation.

In particular, spin-based computational devices built with materials with magnetic order and spin polarized transport have emerged as a viable beyond CMOS option, due to their potential to operate at energy levels close to $1000kT - 100kT$[7]. These devices are a subset of the beyond CMOS devices which include devices based on electron spin[8,9], electron tunneling[10], ferro-electric[11], magneto-electric[12] and multi-ferroic[13] phenomena. Furthermore, these devices also possess favorable attributes such as: 1) non-volatility (ability to retain the information in the absence of power); 2) higher logical efficiency (i.e., fewer devices required to realize a logic function); and 3) high integration density due to these devices being all-metallic and therefore compatible with the state-of-art back-end electronics manufacturing processes. However, spin-based devices have high error rates of 1 in 10 or more precisely due to the need to operate at an energy significantly below the present state-of-the-art electronic switches[14,15]. Hence, in order to design reliable spin-based computing systems there is a need to investigate computational frameworks that can compensate for or tolerate errors. In a sharp contrast to computing systems, communication systems today transmit data reliably i.e., with error rates $p_e < 10^{-12}$ even when the error rate $\epsilon$ of the physical channel is as high as $10^{-2}$. This enormous reduction in error rates was enabled by the foundational work of Shannon[16]. However, Shannon theory has not had much impact in the field of computing even though the potential for such an impact was predicted by von Neumann[17] – an aspiration that remains to be fulfilled to this day. This article we describe a Shannon-inspired approach to design reliable computing systems using high error-rate spin devices.

The problem of designing reliable computing systems using erroneous components was first addressed by von Neumann[17] who defined a reliable logic network as one whose output exhibits a probability of error $p_e < 0.5$ when designed using $\epsilon$-noisy logic gates, i.e., gates whose outputs are in error with probability $\epsilon$. Von Neumann showed that a reliable logic network can be designed for any logic function provided $\epsilon \leq 0.0073$ and that it is impossible to do so if $\epsilon \geq \frac{1}{6}$. Later upper bounds on $\epsilon$ were obtained in a series of papers[18,19] culminating with that of Evans and Schulman[20]. These works proved a precise formula for the upper bound $\epsilon_o$ such that any logic function can be reliably computed using $k$-input $\epsilon$-noisy gates provided $\epsilon < \epsilon_o$. In these works, the design method for constructing reliable logic networks was to replicate individual logic gates and then majority vote the output – the $N$-modular redundancy (NMR) method. NMR is both area and energy expensive. Other fault-tolerant techniques such as check-pointing[21] also incur a significant energy cost. Thus, the problem of designing a reliable and energy efficient computing system using dynamic stochastic switching elements remains an open problem to this day.

In this article, we show that the Shannon-inspired statistical computing (SISC) framework (shown in Fig. 1) enables the design of reliable all-spin logic (ASL) networks operating in stochastic regime (error rate $\epsilon \approx 10\%$) and achieves energy efficiency improvement over an equivalent CMOS implementation. The ASL gate is operated in a stochastic regime – the Shannon ASL (S-ASL) gate – and used as a primitive for constructing Shannon-inspired computing systems. In the language of Shannon theory, in Figure 1, the main computational block is a channel processing information bits while the low-complexity estimator is a side channel transmitting parity bits. The fusion block acts as a decoder that generates its best estimate of the correct output by observing the outputs of the main computational block and the estimator.

As shown in Fig. 1, a high-complexity main block is constructed using high $\epsilon$, low energy, S-ASL gates which result in computational errors $\eta$. To compensate for $\eta$, a low-complexity statistical error compensator comprising an estimator with estimation errors $e$ and a fusion block, is constructed from low $\epsilon$ S-ASL gates. A central concept in Shannon-inspired statistical computing is to transform the network topology of the main block and the estimator algorithm so as to generate a disparity between the probability mass functions (PMFs) $P_\eta(\eta)$ and $P_e(e)$. This disparity is the key to enhancing the robustness of the architecture with minimal overhead error compensation. Our prior work[22] has shown that compensating for errors with low overhead is possible when the error PMF $P_\eta(\eta)$ is *sparse* and $P_e(e)$ is *dense*. Obtaining a dense $P_e(e)$ is not a problem as such PMFs arise in most estimation algorithms. In this article, we show that sparse error PMFs $P_\eta(\eta)$ can be engineered through logic transformations applied to S-ASL networks. Furthermore, we show that NMR techniques to compensate for errors incurs a large overhead (by a factor of $N$) which nullifies the energy efficiency benefits of spintronics.

*A Stochastic Spin-based Switch and a Spin-based Logic Gate*

An inverting/non-inverting S-ASL gate performing a logic inversion with directionality of signal (information) flow is shown in Fig. 1. It comprises two nanomagnets whose magnetic moment directions represent information bits. Each nanomagnet is shared by two spin conduction channels. The logic gate operates via an asymmetric injection of spin current into the spin conduction channel from the nanomagnets. This spin current, in turn, exerts a torque on output magnet forcing it to switch (see methods for the details). The output magnet, however, may not switch with finite probability $\epsilon$ due to the presence of Langevin thermal noise, making a logical error at the gate output (see methods for the details). A complete combinatorial spin logic family including spin majority gates[23], spin interconnects[24,25], spin state elements and random access spintronic memory can be developed and employed to construct a general-purpose Turing-complete computer[26].

We observe that there is a trade-off between gate-level error rate $\epsilon(E, T_g)$, the switching energy $E$, and the delay $T_g$ in S-ASL gates as shown in Fig. 2A. While communication systems comprehend this trade-off between channel error rate and transmit energy, the trade-off between $\epsilon(E, T_g)$, $E$, and $T_g$ is novel for computation [27]. In order to capture this tradeoff, the stochastic behavior of the S-ASL gate, the switching of magnetic moment of the output magnet in particular, is modeled as a Wiener process with a stationary noise term calibrated by fluctuation dissipation theorem[28]. Corresponding Fokker-Planck (FP) equation[29] of the switching magnetic layer (assuming uniaxial anisotropy commonly used in scaled nanomagnetic devices) is solved numerically to capture the time evolution of the probability distribution ($\rho(\theta)$) of the angle $\theta$ made by the magnetization vector with the anisotropy direction. We validated the stochastic properties using extensive time domain Monte Carlo simulations with Stratanovich implementation and FP equations. Fig. 2A shows the energy-delay contours of S-ASL inverter at different error rates $\epsilon$. While competitive with CMOS at $\epsilon = 0.5$, i.e., at average switching delay, S-ASL is not energy competitive at very low error rates of $\epsilon \approx 10^{-15}$ which is needed by traditional von Neumann architectures.

We develop a modified ϵ-noisy gate (Fig. 2B) to describe S-ASL gate which comprehends the underlying physical stochastic behavior, while being sufficiently abstract to permit the design and analysis of complex ASL networks. The modified ϵ-noisy gate comprehends: a) the multiplicative noise in the dynamics of the nanomagnets undergoing field and spin transfer switching, and b) the input dependence of S-ASL errors (the S-ASL gate makes an error only when the output needs to switch and does not, implying a dependence of the error event on the input data). Thermal noise appears as an antipodal ($\pm V$) variable and has a multiplicative effect on the output of a S-ASL gate. It is assumed that Boolean inputs $A_t$ and $B_t$ are provided at time $t$ and the S-ASL gate generates its output $C_{t+T_g}$ at time $t + T_g$ where $T_g$ is the delay *assigned* to the S-ASL gate. The model is composed of an ideal noise-free Boolean gate whose output $C_o = A_t B_t$ is

EXORed with a Bernoulli noise random variable $\theta \in \{0,1\}$ with parameter $\epsilon$, i.e., $\Pr\{\theta = 1\} = \epsilon = 1 - \Pr\{\theta = 0\}$. The output selector computes the final output $C_{t+T_g}$ by choosing either the output of the EXOR gate $C_e = C_o \oplus \theta$ or the error-free output $C_o$. This EXOR gate output is chosen only if $C_o \neq C_t$, where $C_t$ is the S-ASL gate output at time $t$. The $\epsilon$-noisy gate models noise as being additive in the Galois Field of 2 ($GF(2)$) which is equivalent an antipodal multiplicative noise variable $\in \{\pm 1\}$.

*Shannon-inspired Statistical Computing*

The Shannon-inspired statistical computing architecture shown in Fig. 3A includes: 1) a main block that does the bulk of the computation using energy efficient but high error rate circuits, 2) a low-complexity estimator that computes an estimate $y_e$ of the main block output $y_a$, and 3) a low-complexity fusion block that combines where $y_a$ and $y_e$ to generate an error compensated output $\hat{y}$ that is 'close', in a statistical sense, to the correct (but unknown) output $y_o$. Both the estimator and the fusion blocks are designed with low error rate, and hence energy inefficient, circuits. For example, in this article, we design the main block using S-ASL gates with $\epsilon = 0.01$ and the estimator and fusion blocks are designed using S-ASL gates with $\epsilon = 10^{-6}$. The main block and estimator outputs are described by an additive noise model given by

$$y_a = y_o + \eta$$

$$y_e = y_o + e$$

where $\eta$ is the architectural level hardware error caused by the $\epsilon$-noisy S-ASL gates in the main block, and $e$ is the estimation error. Shannon-inspired statistical computing employs the statistics of the input data $x$, the functionality of the main block, and the statistics of $\eta$ and $e$ to obtain the error compensated output $\hat{y}$ efficiently. For example, the knowledge of the precision and functionality of the main block allows one to employ as the estimator a low-precision version of the main block ($m < k$ in Fig. 3A).

As the estimation error $e$ is algorithmic in nature, it is independent of the hardware error $\eta$ whose source is Langevin thermal noise. This independence between the two error sources allows for the realization of a low complexity fusion block which computes a statistical estimate $\hat{y}_o$ of $y_o$ in terms of $y_a$ and $y_e$. Though several methods to realize estimators and the fusion block have been proposed[30], in this work, we consider the method of algorithmic error cancellation (AEC)[22], where the fusion block operates on the difference $y_a - y_e = \eta - e$. As this is a difference between two independent random variables $\eta$ and $e$, the probability mass function of this difference is a convolution of $P_\eta(\eta)$ and $P_e(e)$, the PMFs of $\eta$ and $e$, respectively. The PMF of $e$ is typically dense and centered around zero as shown in Fig. 3B via the central limit theorem. If the PMF of $\eta$ is *sparse* as shown in Fig. 3B then it can be shown[22] that the fusion block can compute an approximation to the *maximum á posteriori* (MAP) rule that chooses $\hat{y}$ to maximize the probability $P(\hat{y} = y_o | y_a, y_e)$.

$$\hat{y} = y_a - 2^{l-s} \left\lfloor \frac{y_a - y_e}{2^{l-s}} + \frac{1}{2} \right\rfloor$$

where $l$ is the precision of the main block output, and $s = \lfloor \log_2 p_k \rfloor$ with $p_k$ denoting the number of distinct peaks in the sparse $\eta$ PMF. This results in a low complexity fusion block architecture as shown in Fig. 3C. The second term in the equation above is a MAP estimate $\hat{\eta}$ of the hardware error $\eta$. Such a fusion block needs three adders and this complexity is independent of the complexity of the main block and the estimator. Thus, the fusion block overhead will be a small fraction of the main block as the latter's complexity increases.

The key to low complexity statistical error compensation is the existence of a low complexity estimator and our ability in ensuring a strong disparity between the PMFs of $\eta$ and $e$. One way, as described above, of ensuring this strong disparity, is to engineer a sparse $P_\eta(\eta)$ ($\eta$ takes few widely separated large magnitude values with high probability) so that it is very different from the $P_e(e)$, which is typically dense (region of support is centered in a small neighborhood around 0)

*Engineering Sparse Error PMFs in Computation*

We show that the requirement of sparse PMFs can be imposed on the outputs of a complex S-ASL network in order to enable SISC techniques. The distribution $P_\eta(\eta)$ is sparse when $\eta$ is restricted to take few large magnitude values with high probability. We show how sparsity (large magnitude $\eta$) is enforced via two logic transformations: 1) inter path delay balancing (I-PDB), and 2) intra path delay redistribution (I-PDR). Figure 4A shows an S-ASL network of a 8-bit ripple carry adder (RCA) which adds two 8-bit inputs $x_1$ and $x_2$ to obtain a 8-bit output $y_a = x_1 + x_2 + \eta$, and an output carry. When all S-ASL gates have the same delay $T_{g,u}$ and energy $E_{g,u}$, and hence, same error rate $\epsilon_{g,u} = \epsilon(E_{g,u}, T_{g,u})$, the resulting error PMF $P_\eta(\eta)$ is dense at $\epsilon_{g,u} = 0.1$. In this case, paths with the largest number of gates $N_{cp}$, i.e., the critical paths, have the highest path delay $T_{cp,u} = N_{cp} T_{g,u}$, which determines the overall throughput. In I-PDB, delays of gates lying on the shorter paths are increased, at a constant energy $E_{g,u}$, such that every gate lies on at least one critical path, i.e., the path with path delay $T_{cp,u}$. Thus, I-PDB reduces error rate of many gates on shorter paths, while it leaves the paths having $N_{cp}$ gates untouched with all their gates operating at high error rate as shown in Fig. 4B. This enhances the error PMF sparsity since the number of gates in a path computing an output bit increases with the significance of the bit in least significant bit (LSB)-first arithmetic architectures such as the RCA. Sparsity can be increased further by using I-PDR where the S-ASL gates delays along each path are redistributed while keeping total path delay, hence the throughput constant, as shown in Fig. 4C. This redistribution is done such that the error rate of the top few MSBs is greater than those of the other bits. In particular, the S-ASL gates at the start and at the end of the critical path are assigned lower delay (higher $\epsilon$) as compared to those that are in the middle. Doing so results in a highly sparse PMF $P_\eta(\eta)$ as indicated in Fig. 4C. Note that all the delay reassignments in both I-PDB and I-PDR techniques are done at a constant switching energy of $E_{g,u}$ (moving vertically in Fig. 2A) and at constant throughput (identical critical path delay $T_{cp,u}$). We define the average device error rate of the architecture as $\epsilon_{cp-avg} = \epsilon(E_{g,u}, T_{cp-avg})$, where $T_{cp-avg} = \left(\frac{T_{cp,u}}{N_{cp}}\right)$. Note: $T_{cp-avg} = T_{g,u}$, when all gates have equal delay. All three architectures and corresponding $\eta$ PMFs (which are obtained for equivalent 15-bit RCAs) in Fig. 4 are compared at $\epsilon_{cp-avg} = 0.1$, same energy and throughput. The supplementary information elaborates general algorithms for I-PDB and I-PDR applicable to any logic network. Thus, the I-PDB and I-PDR are delay manipulation techniques at the gate-level to achieve effective error statistics shaping at the output of a given multi-bit output logic network. However, they can be applied at the level of clusters of logic gates such that all the logic gates in any give clusters can have identical supply current and delay. The size (average gate count) of such clusters can be chosen appropriately to trade-off the design complexity with the effectiveness of error statistics shaping, and hence system-level performance.

*A Support Vector Machine (SVM) classifier using S-ASL Gates*

We demonstrate the benefits of SISC for a S-ASL implementation of SVM classifier used for electroencephalogram (EEG) based seizure detection. The SVM implementation is presented with input data $x \in \mathbb{R}^N$, which is $N(=120)$ dimensional feature vector extracted from the EEG signal. It assigns a label $\hat{z} \in \{-1,1\}$, where $\hat{z} = 1$ indicates the presence of seizure. The label $\hat{z}$ is calculated according to the decision rule $\hat{z} = \text{sign}(w^T x + b)$ where $w \in \mathbb{R}^N$ and $b \in \mathbb{R}$ are the $N$-dimensional weight vector and a scalar bias, respectively. Let $z \in \{-1,1\}$ denote the true label for the input vector $x$. The accuracy of the classifier implementation is captured in terms of true positive (TP) rate $p_{TP}$ and false alarm (FA) rate $p_{FA}$, where $p_{TP} = \Pr\{\hat{z} = 1 | z = 1\}$ and $p_{FA} = \Pr\{\hat{z} = 1 | z = 0\}$, and the probabilities are estimated empirically (via leave-one-out cross-validation[31]) for the MIT-CHB EEG dataset[32] by running extensive Monte Carlo simulations.

Figure 5A shows the conventional serial architecture employing $N$ Baugh Wooley multipliers (BWM) and a carry save adder (CSA). All gates in this architecture operate at identical error rates. The Shannon-inspired architecture in Fig. 5B employs the serial architecture as the main block (MB), and applies I-PDB and I-PDR to it in order to shape its output error distribution. Since I-PDB and I-PDR techniques make some gates operate at lower error rate, few reliable intermediate signals in BWMs can be employed as the estimates of their outputs indicated via green RPE-EST blocks in BWMs (see methods for the details). We add an error compensation (EC) block consisting of a CSA and a fusion block computing final output $\hat{z}$. The overhead of EC block amounts to ~11% of the gate complexity of the MB. In the EC, we assume same low error rate $\left(\epsilon \approx 10^{-4} \epsilon_{cp-avg}\right)$ for all the gates in EC to keep its design simple. We assume that the fusion block computation can be pipelined since it operates only on the final outputs of the MB and the estimator. This allows the gates in fusion block to operate at much lower energy since its critical path is much shorter than that of the main block.

We show that SISC outperforms the traditional serial architectures and $N_m$ modular redundancy ($N_m$-MR) architectures, which replicate the conventional serial architecture $N_m$ times and take bitwise majority vote on their outputs. We compare the $p_{TP}$ of the serial architecture (figure S9), the Shannon-inspired architecture (figure S10) and a 3-MR architecture at a constant $p_{FA} = 1\%$. We observe in Fig. 6A that Shannon-inspired architecture can tolerate $1000 \times$ higher device error rate ($\epsilon_{cp-avg}$) compared to the serial architecture while maintaining the $p_{TP}$ close to that of the fixed point ideal error-free implementation. In particular, the TP rate for Shannon-inspired architecture is close to 93% even though the device error rate $\epsilon$ is as high as 1%. The 3-MR architecture tolerates a device error rate (up to 0.01%) that is greater than that of the serial architecture but lower by $100 \times$ when compared to the Shannon-inspired architecture. Conventional 20nm LV CMOS based von Neumann architecture is designed to operate very reliably ($\epsilon = 10^{-15}$) by carefully budgeting the impact of process, voltage and temperature variations. This Shannon-inspired approach demonstrates $10^{13}$ fold increase in tolerable device error rates, while maintaining the system-level performance.

The ability of Shannon-inspired architecture to tolerate high device error rate translates into gains in energy-efficiency when compared to both serial and 3-MR architectures. We compare total energy/decision for fixed decision delay of 9.7155ns. The Shannon-inspired architecture achieves $3 \times$ lower energy compared to the serial architecture (Fig. 6B) while maintaining $p_{TP} = 93\%$, thanks to its $1000 \times$ higher device error rate tolerance. The 3-MR architecture, however, consumes $2.3 \times$ *more* energy even after tolerating marginally higher device error rate. This is because the energy overhead of replication overweighs the energy reduction achieved via tolerating higher device error rate. Thus, the Shannon-inspired computing outperforms the conventional architectures in both energy and accuracy with substantial margins. This improvement in energy makes the Shannon-inspired architecture competitive to CMOS. All the reported energy numbers here do not include the leakage energy or the energy consumed in the interconnects. For

S-ASL based implementations, we do not include the energy consumed in clocking network for all three architectures.

We have demonstrated the benefits of designing reliable inference systems on stochastic spintronic devices by leveraging Shannon theory for error compensation. The ability to perform reliable computation on stochastic device fabrics can enable the use of a highly error prone but scalable physical devices.

## Methods:

**Description of spin torque logic device, its operation, and estimation of its nominal delay and switching energy:**

As an example spintronic logic device amenable to Shannon inspired computing, we describe a representative nanoscale logic device, referred to as the All Spin Logic (ASL) device, which operates via generation and interaction of spin currents with nanomagnets. The intrinsic ASL device consists of two nanomagnets (with magnetization pointing out of plane) sharing a spin conduction channel as shown in Extended Data (ED) Fig. 1A. The information bit is represented by the direction of the magnetization vector of the nanomagnet. Each nanomagnetic node connects to two spin conduction channels for a) receiving spin information b) regenerating the signal information. For positive supply voltages, the intrinsic ASL device operates as a Boolean inverter with directionality of signal (information) flow is shown in ED Fig. 1A. The input and output nanomagnets inject electrons into the spin conduction channel, where the electron spin direction is decided by the orientation of the nanomagnets (for example, nanomagnet with spin pointing to the **+X** direction injects electrons oriented in the **-X** direction). The dominant magnet, therefore, sets up a larger spin polarization with orientation opposite to its own magnetic moment. In ED Fig. 1A and 1B, the nanomanget M1 is made dominant by allocating larger overlap area with the spin channel compared to M2. A spin current flows in the channel from the higher spin potential[8] (M1) to the nanomagnet M2 creating a spin torque to orient its magnetic moment in direction opposite to that of the M1. For negative supply voltage, a converse process happens forcing M2 to align parallel to M1 and the device acts as buffer. The ASL device is also amenable to integration into a microchip (ED Fig. 1B). Majority gates can be created by having more input magnets sharing the conduction channel.

The example materials for forming the nanomagnets can be CoFeB, CoFe-based Heusler alloys, $Mn_xGa_y$ class of materials, or L10 metals (FePt, FePd). Heusler alloys having a large range of magnetic anisotropy ($H_k$) and low saturation magnetization ($M_s$) are particularly suitable[9]. Example channel materials are $Si^{24}$, $Cu^{25}$ and Graphene[33] which exhibit excellent room temperature

We evaluate the energy-delay product of nominal the ASL device (comprising magnets with 52kT energy barrier corresponding to more than 7-year retention time for 1 bit) with nominal existing device parameter choices[9] (ED Table 1). Materials with existing proof of concept integration in CMOS are chosen. At nominal operating voltage of 10 mV, the ASL device operates with a response time ~0.5 ns, which is dominated by the delay of switching. The estimated energy-delay product of the device is given by the total joule energy supplied by the supply voltage. An example time domain switching dynamics of ASL device is shown in ED Fig. 1C, while corresponding spin current magnitudes are shown in ED Fig. 1D, and instantaneous power consumption in ED Fig. 1E. These plots are obtained by performing simulations of SPICE-based models[34] of ASL. Exemplary equivalent circuit[34] is shown in ED Fig. 1F.

Since the nanomagnets are non-volatile, the information bit is preserved even though the power supply of ASL gate is turned off after its operation. It has already been shown[23] that it is energy-efficient to clock the ASL gates such that they are turned ON only when they need to process the information. Such clocking scheme eliminates the static power consumption to large extent. Here we assume that the ASL gates are clocked. Hence, the delay of any ASL gate is determined by the time duration for which the gate is turned ON. We denote gate delay by $T_g$.

The energy consumption in the ASL based Boolean inverter is given by

$$E_{\text{inv}} = I_{\text{supply}}^2 R_{\text{spin}} T_g \tag{A1}$$

where $I_{\text{supply}}$ denotes the charge current supplied to the magnet and $R_{\text{spin}}$ is the series electrical resistance of the magnet and the channel. The energy consumption of 3-majority gate is given by

$$E_{\text{MAJ3}} = 3 I_{\text{supply}}^2 R_{\text{spin}} T_g$$

since current $I_{\text{supply}}$ needs to be passed through three identical magnets to operate a 3-majority gate.

**Estimation of Logic Error Rates using Fokker-Planck equation**

The response time of ASL device is dominated by the switching time of output nanomanget, which is a random variable due to the randomness in the initial direction of the magnetic moment when the switching process starts. The phenomenological equation describing the dynamics of nanomagnet with a magnetic moment unit vector ($\hat{m}$), the modified Landau-Lifshitz-Gilbert (LLG) equation, is (see ED Table 1 for parameters)

$$\frac{\partial \hat{m}}{\partial t} = -\gamma \mu_o [\hat{m} \times \overline{H}_{\text{eff}}] + \alpha \left[ \hat{m} \times \frac{\partial \hat{m}}{\partial t} \right] + \frac{\vec{I}_\perp}{eN_s} \tag{A2}$$

where $\gamma$ is the electron gyromagnetic ratio, $\mu_0$ is the free space permeability, $\overline{H}_{\text{eff}}$ is the effective magnetic field due to material/geometric/surface anisotropy, $\alpha$ is the Gilbert damping of the material, and $\vec{I}_\perp$ is the component of vector spin current perpendicular to the magnetization ($\hat{m}$) leaving the nanomagnet, $N_s$ is the total number of Bohr magnetons per magnet. It has been proposed that stochasticity in the initial direction can be equivalently modeled as additive random noise field[15]. The noise field acts isotopically on the magnet. The internal field is described as:

$$\overline{H}_{\text{eff}} = \overline{H}_{\text{eff,m}} + H_{n,i}\hat{x} + H_{n,j}\hat{y} + H_{n,k}\hat{z}$$

Where $\overline{H}_{\text{eff},m}$ denotes the mean effective magnetic field due to material/geometric/surface anisotropy, while, the first and second order moments of random noise field components $H_{n,i}, H_{n,j}$ and $H_{n,k}$ as a function of time are given as

$$\langle H_{n,l}(t)\rangle = 0$$

$$\langle H_{n,l}(t)H_{n,m}(t')\rangle = \frac{2\alpha k_B T}{\mu_o^2 \gamma M_s V}\delta(t-t')\delta_{lm}$$

for $m, l \in \{i, j, k\}$, where $M_s$, $V$ denote saturation magnetization and volume of the nanomagnet, respectively, and $k_B$ and $T$ denote Boltzmann constant and temperature, respectively. The initial conditions of the magnets should also be randomized to be consistent with the distribution of initial angles of magnet moments in a large collection of magnets. We used a mid-point integration method[35] to apply the Stratonovich calculus while integrating the LLG equation to compute orientation of $\hat{m}$ as a function of time. The ASL device is said to have made a *switching error* if the orientation of $\hat{m}$ does not change appropriately within the time duration $T_g$ even though the charge current $I_{\text{supply}}$ is passed through the magnet to achieve the switching. Such switching errors at the device-level cause logic errors in the computation. One can empirically compute the probability of switching error ($\epsilon$) via Monte-carlo simulations of the magnetization by numerical integration of Landau-Lifshitz-Gilbert (LLG) equation and randomly sampling the noise field. Alternatively, here we use a dynamic equation governing the time domain evolution of the probability of the direction of the magnetic moment, referred to as the Fokker-Planck equation. It solves for the probability distribution of the direction of the magnetic moment[15],

$$\frac{\partial \rho(\theta,\tau)}{\partial t} = -\nabla \cdot J(\theta,\tau) = -\frac{1}{\sin\theta}\frac{\partial}{\partial \theta}[\sin\theta \, J_\theta(\theta,\tau)] \tag{A3}$$

where $\rho(\theta,\tau)$ is probability density of $\hat{m}$ (with angle and time as variables, see ED Fig. 2A & 2B, for example) and $J_\theta(\theta,\tau)$ is the flow of probability given by drift and diffusion components

$$J_\theta(\theta,\tau) = \rho(\theta,\tau)\frac{\partial \theta}{\partial \tau} - D\frac{\partial \rho(\theta,\tau)}{\partial \tau} \tag{A4}$$

The flow term $\frac{\partial \theta}{\partial \tau} = (i - h - \cos\theta)\sin\theta$ where $i = \frac{I_{\text{supply}}}{I_{\text{crit}}}$, $h = \frac{H}{H_{kc}}$ are the current and field driving terms for the probability, $D = \frac{kT}{2E_b}$ is the diffusion constant in terms of the thermal barrier $E_b$ of the magnet, and $I_{\text{crit}}$ denotes the minimum current required to switch of the nanomagnet, referred to as the critical current of the nanomagnet. Also,

$$\tau = \frac{\alpha\gamma\mu_0 H_{\text{eff}}}{1+\alpha^2}t.$$

where, $t$ denotes time. We have compared the Fokker-Planck models with Monte-carlo simulations of the magnetization by numerical integration of Landau-Lifshitz-Gilbert (LLG) equation (see validation in Fig. 2C). The logic error rate ($\epsilon$) at any time $\tau$ can be obtained by integrating $\rho(\theta, \tau)$ with respect to $\theta$ varying from 0 to $\frac{\pi}{2}$, assuming the initial orientation of magnetic moment $\hat{m}$ is $\theta = 0$.

One can derive approximate analytical expression for error rate $\epsilon$ as[14]

$$\epsilon(i, T_g) = 1 - \exp\left[\frac{-\pi^2(i-1)\frac{E_b}{4kT}}{ie^{\frac{2\alpha\gamma H_k \mu_0 T_g(i-1)}{(1+\alpha^2)}} - 1}\right] \qquad (A5)$$

by assuming that the delay of ASL gate is dominated by the time required for the output nanomanget to switch.

If $i \gg 1$, the $\epsilon(i, T_g)$ expression can be approximated as,

$$\epsilon(i, T_g) \approx 1 - \exp\left[\frac{-\pi^2 \frac{E_b}{4kT}}{e^{\frac{2\alpha\gamma H_k \mu_0 T_g i}{(1+\alpha^2)}}}\right]$$

Denoting the energy delay product of the gate as $K_g$, we have,

$$E_g T_g = K_g$$

and using equation (A1), we get,

$$\epsilon(K_{g_i}) \approx 1 - \exp\left[\frac{-\pi^2 \frac{E_b}{4kT}}{e^{\frac{2\alpha\gamma H_k \mu_0 \sqrt{K_g/R_{\text{spin}} I_{\text{crit}}^2}}{(1+\alpha^2)}}}\right] \qquad (A6)$$

Thus, the device error rate $\epsilon$ remains approximately constant for constant energy delay product $K_g$ for a given gate $g$. This is the reason why iso-$\epsilon$ contours in Fig. 2A appear approximately as straight lines. For a given device, its error rate can be reduced by increasing its energy-delay product either via increasing its current or increasing its delay or some combination of the two. This insight is exploited by logic transformation techniques I-PDB and I-PDR to achieve error statistics shaping as described in the main text.

Above analysis is for an ASL-based Boolean inverter. For 3-majority gate, the error rate corresponds to either supply current of $3I_{\text{supply}}$ or $I_{\text{supply}}$ depending on whether all three input nanomagnets have parallel magnetic moments or not. However, we here conservatively assume a constant error rate corresponding to $I_{\text{supply}}$ since such input dependence of error rate is too complex to tract in Monte Carlo simulations.

**Design of Support Vector Machine (SVM) classifier**

Here we elaborate some of the techniques applied in the Shannon-inspired architecture of the SVM implementation.

**Constrained I-PDB:** As described in main text, in I-PDB, the gates on the non-critical paths are made to operate at correspondingly larger delay, but at constant energy, and hence reducing their error rate. However, if the main block is sufficiently complex, it is possible that few primary paths, being very short compared to the critical paths, allow sufficiently large increase in the gate delay. In inference applications, reducing the error rate below $10^{-6}$-$10^{-7}$ may not lead to any further robustness benefits (since, the feature noise starts dominating the final system-level performance). In such cases, one can constrain the I-PDB technique to increase the gate delay, while decreasing its switching energy and maintaining its error rate sufficiently low. The impact of constrained I-PDB is illustrated in ED Fig. 3A. Suppose given gate $g_1$ is initially at $(E_{g1}, T_{g1}, \epsilon_{g_1})$ and application of I-PDB results in the delay of $T^*_{g1}$ and reduction in its error rate to $\epsilon^*_{g_1}$. Now, if corresponding inference kernel implementation preserves its system-level performance even if all its gates have error rates of $\epsilon'_{g_1}$, the I-PDB can be constrained to increase its delay to $T^*_{g1}$ while maintaining its error rate to $\epsilon'_{g_1}$ and reducing its energy to $E'_{g1}$ $(< E_{g_1})$. Thus the constrained I-PDB can achieve more energy-efficient design compared to the I-PDB, while maintaining the system-level performance.

**Reduced-precision embedded estimator (RPE-EST):** Consider a multiplication of two 8-bit binary operands, $a_8 a_7 \ldots a_1$ and $b_8 b_7 \ldots b_1$, where $a_i, b_i \in \{0,1\} \forall i \in \{1, \ldots, 8\}$ as follows:

$$m_o = [c_{16} \ldots c_2 \, a_1 b_1] = (a_8 a_7 \ldots a_1) \times (b_8 b_7 \ldots b_1)$$

where $m_o$ is 16-bit binary output with its individual bits denoted as $c_{16}, \ldots, c_2 \in \{0,1\}$. The corresponding architecture of Baugh-Wooley multiplier is shown in ED Fig. 3B, where $p_{ij} = a_i b_j$, $\bar{p}_{i8} = a_i \bar{b}_8$ and $\bar{p}_{8i} = \bar{a}_8 b_i \;\; \forall i,j \in \{1, \ldots, 7\}$. If the input binary numbers are quantized to 5-bits, we get,

$$m_{o5} = [d_{10} \ldots d_2 \, a_5 b_5] = (a_8 a_7 a_6 a_5 a_4) \times (b_8 b_7 b_6 b_5 b_4)$$

where $m_{o5}$ is 10-bit binary output with its individual bits denoted as $d_{10}, \ldots, d_2 \in \{0,1\}$.

Now, we use $m_{o5}$ as an estimate of $m_o$. However, instead of adding a separate 5-bit multiplier, we observe that $m_{o5}$ can be obtained by taking intermediate signals from $8 \times 8$ multiplier architecture as shown in ED Fig. 3B. We need to add four more full adders (shown by dotted green lines in ED Fig. 3B) in order to compute appropriate carry ripple. Thus, the estimate of actual output $m_o$ is obtained at minimal overhead. In addition, since most of the green-colored full adders are on the non-critical paths, the error in $m_{o5}$ is dominated by quantization error having a dense distribution over a shorter range, as desired for effective error compensation. The output $m_{o5}$ obtained this way is exact if both $(a_8 a_7 \ldots a_1)$ and $(b_8 b_7 \ldots b_1)$ are non-negative.

In this SVM implementation, the feature vector $\boldsymbol{x}$ is unsigned while the weight vector $\boldsymbol{w}$ is signed. Hence, an appropriate constant (independent of component magnitudes of vector $\boldsymbol{w}$) is added in the final estimator output via the additional CSA, which adds individual estimates of BWM outputs in the EC block.

**Current redistribution in CSA and dimension reordering in the MB:** Since the effectiveness of the I-PDB and I-PDR techniques increases with increasing path delay diversity in the logic network, we choose serial CSA in this design. We reorder the dimensions of weight vector $\boldsymbol{w}$ such that the products $w_i x_i$ are added in the ascending order of the value of the dimension $w_i$, i.e. $w_j x_j$ is added earlier than $w_i x_i$ if $w_j <$

$w_i$. This enables I-PDB to compute the products of larger dimensions of $w$ with more reliability (since corresponding primary paths are shorter and hence consist of gates operating at lower error rate). Note that this dimension reordering needs to be performed only once. In serial CSA, since all paths have very similar path delays (path delay difference only arises due to the RCA at the last stage), we apply supply current redistribution to the CSA in the MB. In particular, we lower the supply current for gates computing the MSBs while increasing it for the gates computing LSBs in the CSA. It is to be noted that, unlike delay redistribution, there is no constraint on increasing the supply current in the current redistribution. This effectively enables increase in the switching energy of gates in the paths computing LSBs in the CSA until sufficiently good error distribution sparsity is achieved.

**Simulation methodology for S-ASL based architectures and systems:**

We carry out simulations of Fokker-Planck equations to estimate $\epsilon(E, T_g)$ for different values of $E$ and $T_g$ for an ASL-based Boolean inverter, and generate iso-$\epsilon$ contours in Fig. 2A using MATLAB. Once this data is obtained, all subsequent architecture and system level simulations are carried out in MATLAB. We then separately estimate the normalized delay factors after applying I-PDB and I-PDR (as described in supplementary information) for different blocks, such as Ripple Carry Adder (RCA), Baugh-Wooley multiplier (BWM), Carry Save Adder (CSA). For example, the normalized delay factors for RCA (architecture in ED Fig. 4) are shown in ED Table 2, where the delay factor of 1 corresponds to the error rate of 10%. We then design S-ASL based Shannon-inspired architecture implementing SVM using the BWM and CSA as shown in Fig. 5 and apply I-PDB and I-PDR techniques as described earlier in this section, and in supplementary information. We thus obtain ($\epsilon$, $T_g$, $E$) triplet for every gate in the architecture. Then, we carry out Monte Carlo simulations by passing extracted input feature vectors through the architecture (with each gate making error randomly with corresponding $\epsilon$ probability) to estimate the final system-level performance in terms of true positive rate $p_{TP}$ and false alarm rate $p_{FA}$. The decision delay is the sum of gate delays on the critical path, while total energy/decision is the sum of switching energies of all gates. Similarly, we estimate the system-level performance of the serial architecture, where all gates have identical delay and error rate. The $p_{TP}$ and $p_{FA}$ for 20nm LV CMOS are same as that of the error-free computation. We follow the benchmarking methodology[36] to estimate the energy and delay of the CMOS implementation and consider energy-delay curve of 20nm LV CMOS FO4 inverter[9] as shown in Fig. 2A. We assume the activity factor of 50% for CMOS implementation.

**Acknowledgments:** This work was supported in part by Systems on Nanoscale Information fabriCs (SONIC), one of the six SRC STARnet Centers, sponsored by MARCO and DARPA.

**Figure 1**: Shannon all-spin logic (S-ASL) gates trade-off the switching energy $E$ with delay $T_g$ and gate-level error rate $\epsilon$. A Shannon-inspired computing system is able to operate with very low system-level error rates even when the bulk of the computation (main block) is designed using high $\epsilon$ (low $E$) S-ASL gates. We shape the architectural level error distribution at the output of the main block using logic transformations and the $E$ vs. $T_g$ vs. $\epsilon$ trade-off in S-ASL gates so as to enable a low-complexity statistical error compensator (an estimator and a fusion block) to be designed. The error compensator is able reduce the system-level error rate by more than 2-orders-of-magnitude by correcting the computational errors at the main block output. In doing so, substantial gains in both energy consumption and system accuracy are obtained over conventional ASL implementations.

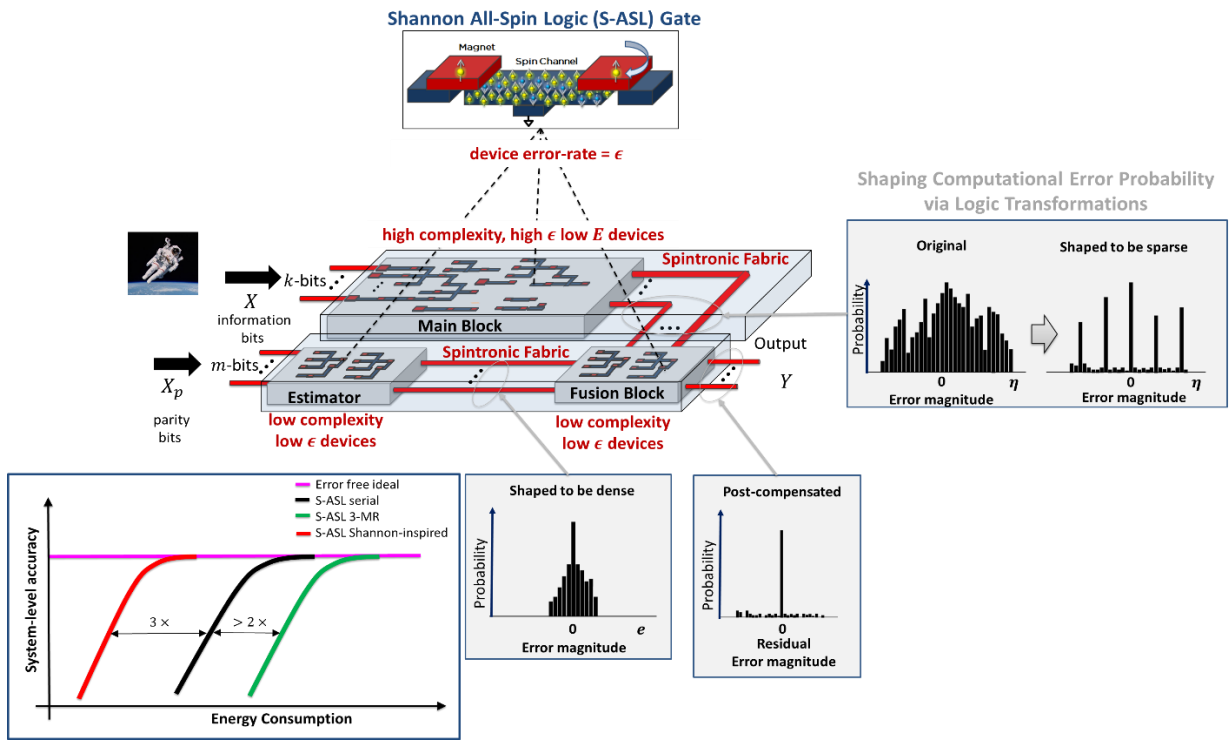

**Figure 2**: The Shannon ASL (S-ASL) gate: (A) a plot showing the trade-off between the gate error rate $\epsilon$, the switching energy $E$, and the delay $T_g$. The energy gap between an (error-free) ASL gate and a S-ASL gate with $\epsilon = 0.4$ is approximately $40 \times$. Error free ($\epsilon < 10^{-14}$) ASL gates needed by the von Neuman architecture can be obtained from S-ASL gates operating at $\epsilon = 0.4$ by either increasing the switching energy by $40 \times$ (keeping delay fixed) or by increasing the delay by $100 \times$ (keeping the switching energy fixed) or some combination of the two. The proposed Shannon-inspired statistical computing framework enables operation using S-ASL gates with $\epsilon = 0.4$. (B) An $\epsilon$-noisy gate model of an S-ASL AND gate showing it as being composed of an error-free AND gate followed by a virtual gate that emulates the stochastic behavior of the spin device.

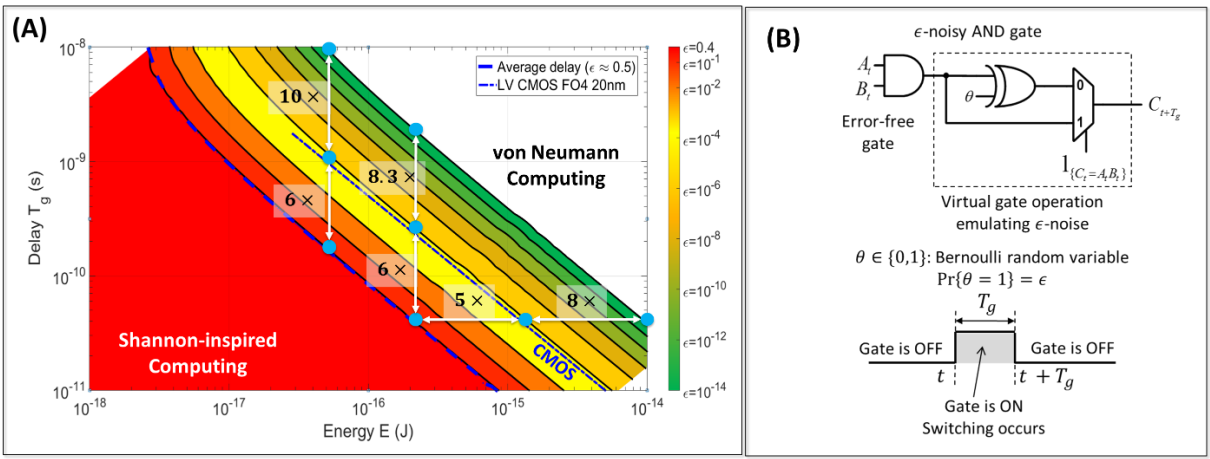

**Figure 3**: Shannon-inspired statistical computing: (A) an architecture showing statistical error compensation, (B) an illustration of the disparity between the PMFs of $\eta$ and $e$ necessary for low complexity error compensation, and (C) the architecture of the fusion block.

**Figure 4**: Engineering the distribution of error in a ripple carry adder (RCA) constructed from S-ASL gates: (A) with uniform delay assignment, (B) with inter-path delay balancing (I-PDB) technique that increases the delay of gates in the non-critical path (at constant energy), (C) with intra-path delay redistribution (I-PDR) technique to reassign gate delays along the critical path (at constant energy). In each subfigure, the color of the box corresponds to a particular error rate regime in Fig. 2A, in which the gate is operating. All three architecture consume identical switching energy and operate at identical throughput corresponding to average device error rate $\epsilon_{\text{cp-avg}}$ of 10%. The $\eta$ distributions are obtained for equivalent 15-bit RCA architecture, and corresponding actual delay factors are given in ED Table 2.

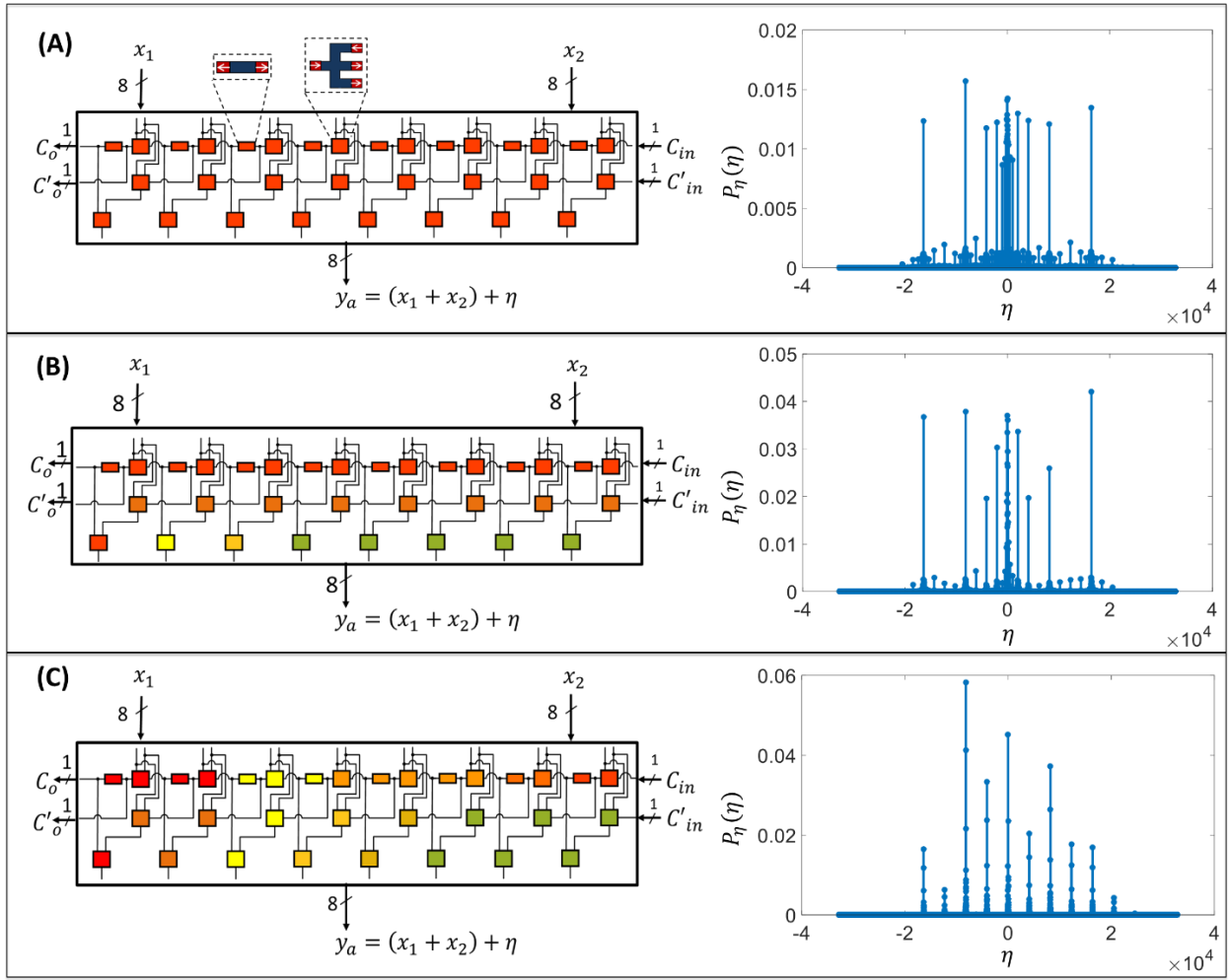

**Figure 5**: The spin-based support vector machine (SVM) classifier using S-ASL gates: (A) the conventional serial architecture with uniform delay assignment, and (B) a Shannon-inspired architecture.

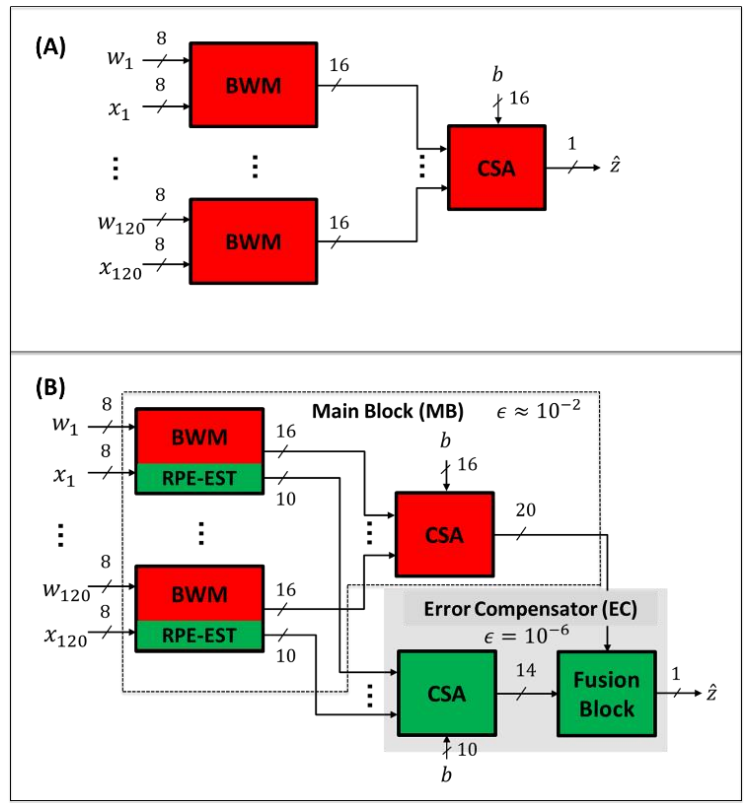

**Figure 6**: True positive (TP) rate of the spin-based support vector machine (SVM) classifier wrt.: (A) $\epsilon_{\text{cp-avg}}$, which is the average device error rate, and (B) total energy consumption per decision at constant false alarm rate of 1%.

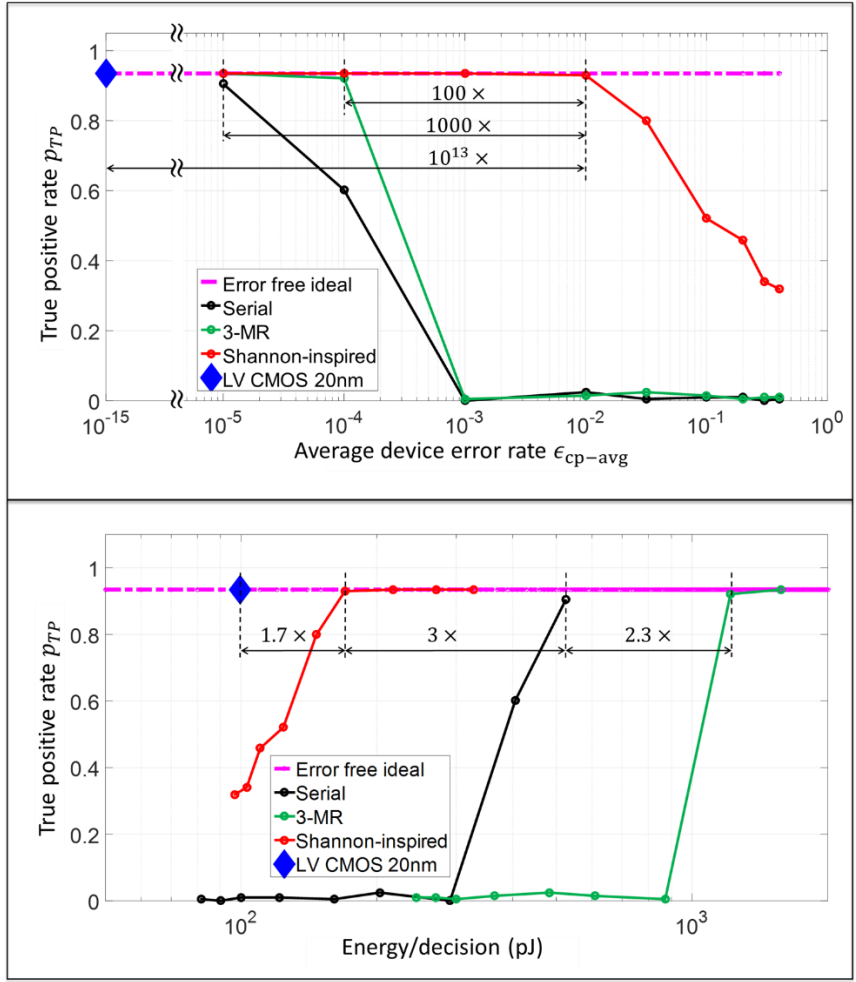

# Extended Data: Figures and Tables

**Extended Data Figure 1**: All Spin Logic (ASL) inverter: (A) A schematic consisting of two nanomagnets interacting via spin channels, (B) cross section schematic, (C) switching of input and output magnetization at supply voltage of 10mV, (D) corresponding spin currents during the switching operation, (E) instantaneous power consumption during the switching operation, and (F) equivalent RC circuit.

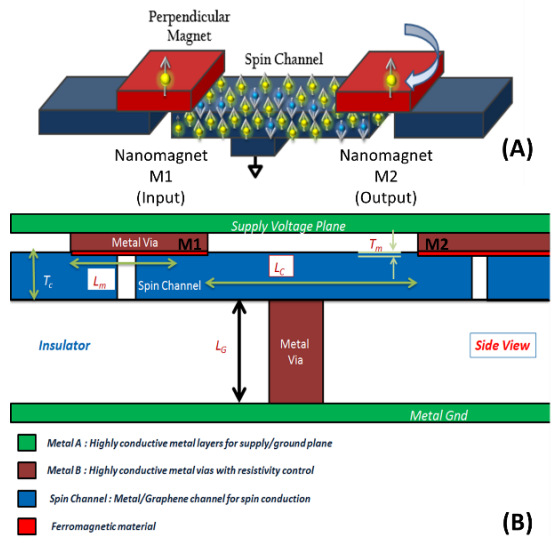
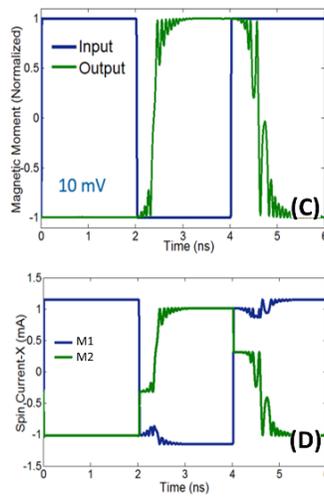
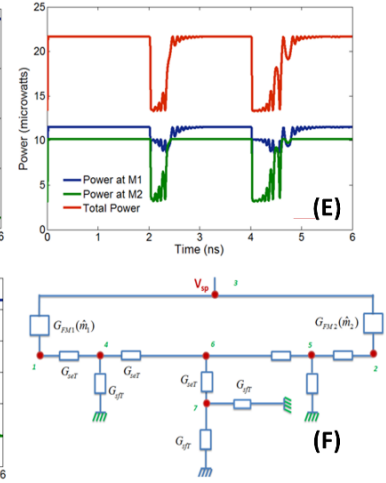

**Extended Data Figure 2**: Time domain evolution of probability distribution $p(\theta, t)$ of magnetic moment of output magnet, and corresponding error rates ($\epsilon$) observed using Fokker-Planck (FP) equation : (A) Probability function evolution for $E_b = 40kT$, for time $t \leq 10$ns, (B) Probability function evolution for $E_b = 20kT$, for time $t \leq 0.1$ ns, and (C) Comparison of the $\epsilon$ for various values of $E_b$ obtained using direct integration of the stochastic LLG equation and FP equation.

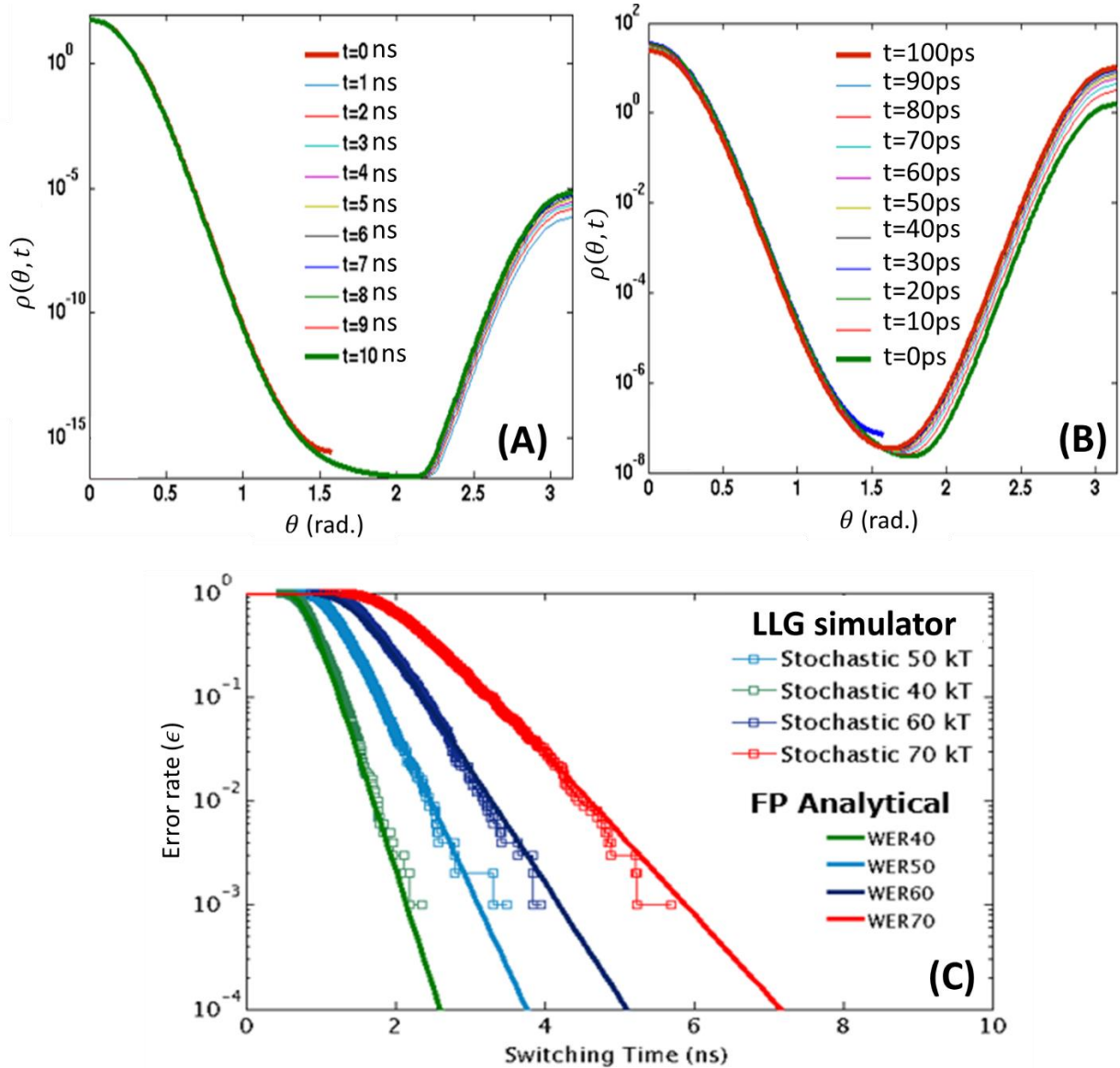

**Extended Data Table 1**: Device parameters considered in the simulations.

| Parameter | Notation | Value | Units |
|---|---|---|---|
| Energy Barrier | $E_b$ | $52kT$ | J |
| Effective Internal Anisotropic field | $H_k$ | $16 \times 10^4$ | A/m |
| Saturation magnetization of magnet | $M_s$ | $250 \times 10^3$ | A/m |
| Damping constant of the Magnet | $\alpha$ | 0.007 | - |
| Resistance×Area for intrinsic spin device | $RA$ | $0.6 \times 10^{-14}$ | $\Omega m^2$ |
| Temperature | $T$ | 300 | K |
| Width of magnet | $W_m$ | 37.8 | nm |
| Length of magnet | $L_m$ | 75.7 | nm |
| Thickness of magnet | $T_m$ | 3 | nm |
| Polarization factor | $P$ | 0.8 | - |
| Length of channel | $L_c$ | 100 | nm |
| Thickness of channel | $T_c$ | 200 | nm |
| Thickness of ground lead | $T_g$ | 100 | nm |
| Length of ground lead | $L_g$ | 200 | nm |

**Extended Data Figure 3**: Details about the design of support vector machine (SVM) implementation: (A) Illustration of the difference between the I-PDB and the constrained I-PDB, and (B) Architecture of $8 \times 8$ Baugh-Wooley multiplier, where each $p_{ij}$ denotes a partial product $a_i b_j$, and $a_i$ and $b_i$ denote the bits of 8-bit binary operands. The green shaded area is a $5 \times 5$ Baugh-Wooley multiplier operating on first 5 input MSBs (after including 4 additional full adder blocks shown in green dotted line). The output $[d_{10}\ d_9\ d_8\ d_7\ d_6\ d_5\ d_4\ d_3\ d_2\ a_5 b_5]$ is the output of RPE-EST block as denoted in the main text Fig. 5.

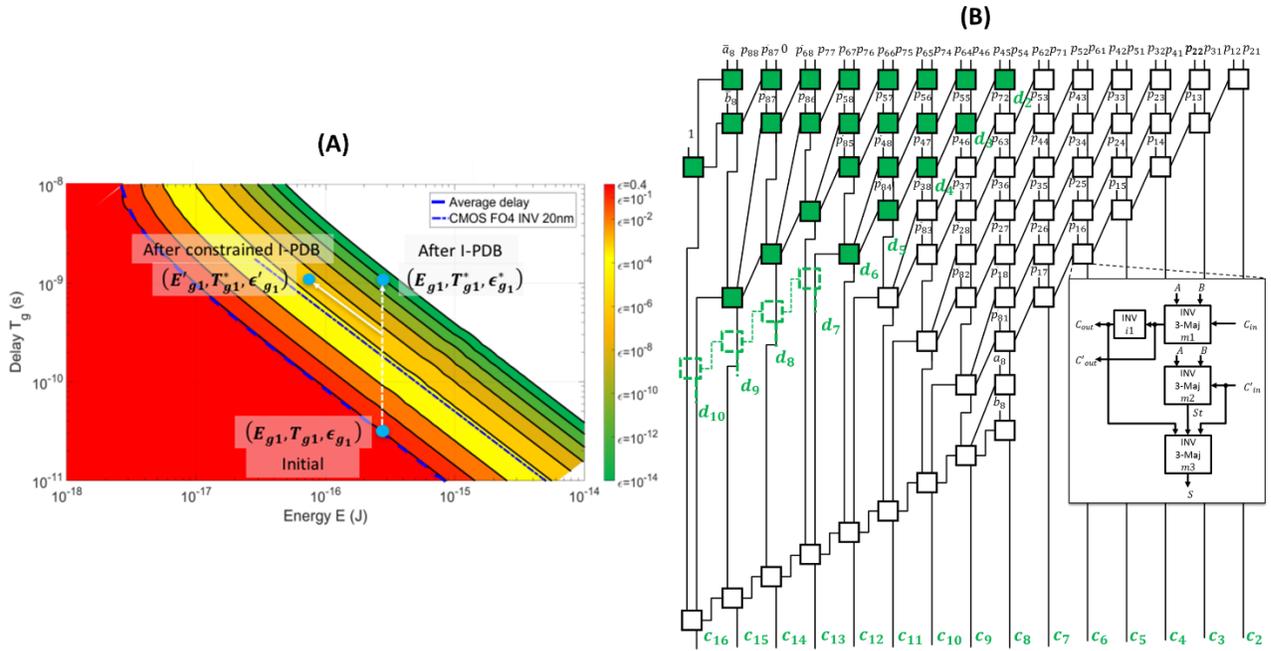

**Extended Data Figure 4**: Architecture of ripple carry adder (RCA).

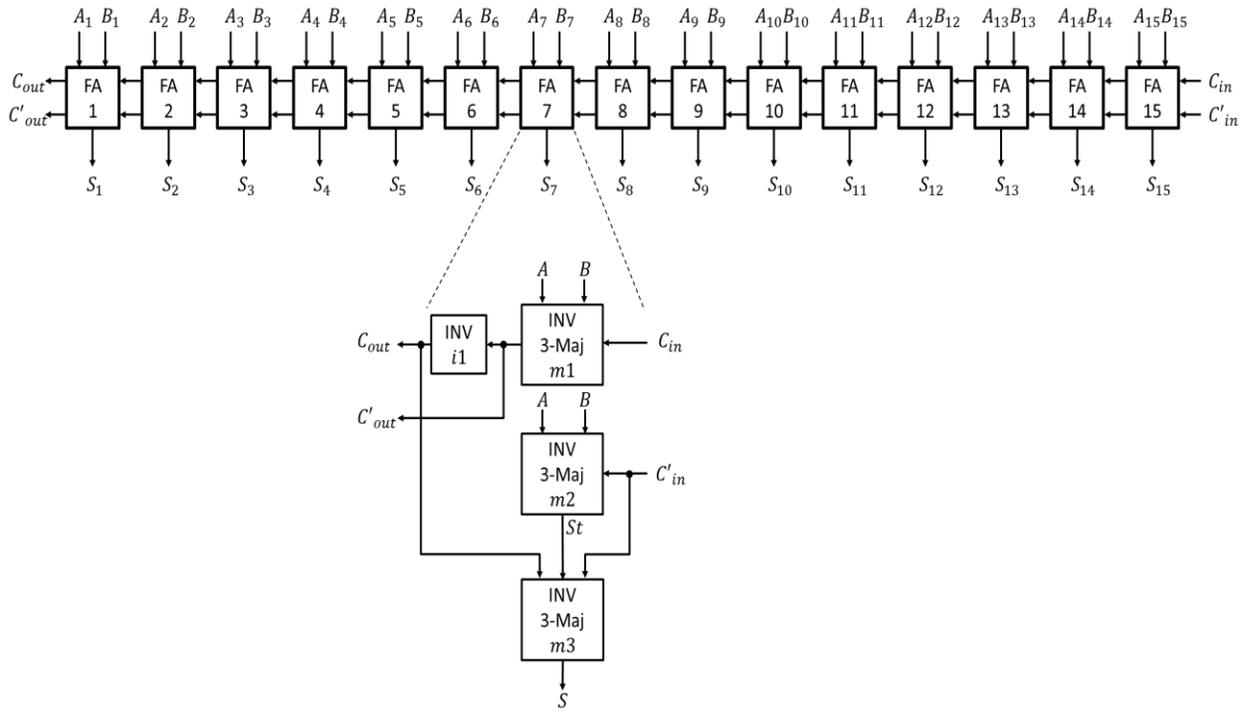

**Extended Data Table 2**: The normalized delay factors for all gates in RCA (indexed as shown in ED Fig. 4) after (A) I-PDB, and (B) I-PDB followed by I-PDR. Here, normalized delay factor of 1 corresponds to error rate of 10%, and all majority gates and inverters consume same energy $E_{maj}$ and $E_{inv}$, respectively, with $E_{maj} = 3E_{inv}$. These relative delay numbers were used to obtain distributions in main text Fig. 4.

(A)

| FA number | gate m3 | Gate m2 | Gate m1 | Gate i1 |
|---|---|---|---|---|
| 1 MSB | 1 | 3 | 1 | 1 |
| 2 | 3 | 3 | 1 | 1 |
| 3 | 5 | 3 | 1 | 1 |
| 4 | 7 | 3 | 1 | 1 |
| 5 | 9 | 3 | 1 | 1 |
| 6 | 11 | 3 | 1 | 1 |
| 7 | 13 | 3 | 1 | 1 |
| 8 | 15 | 3 | 1 | 1 |
| 9 | 17 | 3 | 1 | 1 |
| 10 | 19 | 3 | 1 | 1 |
| 11 | 21 | 3 | 1 | 1 |
| 12 | 23 | 3 | 1 | 1 |
| 13 | 25 | 3 | 1 | 1 |
| 14 | 27 | 3 | 1 | 1 |
| 15 LSB | 28 | 3 | 1 | 1 |

(B)

| FA number | gate m3 | Gate m2 | Gate m1 | Gate i1 |
|---|---|---|---|---|
| 1 MSB | 0.671428571428572 | 2.01428571428571 | 0.671428571428572 | 0.671428571428572 |
| 2 | 2.01428571428571 | 2.01428571428571 | 0.671428571428572 | 0.671428571428572 |
| 3 | 3.10127728174603 | 3.10127728174603 | 0.671428571428572 | 0.671428571428572 |
| 4 | 4.55078125000000 | 4.55078125000000 | 1.46718750000000 | 1.50255456349206 |
| 5 | 5.92955109126984 | 5.92955109126984 | 1.39645337301587 | 1.43182043650794 |
| 6 | 7.23758680555556 | 7.23758680555556 | 1.32571924603175 | 1.36108630952381 |
| 7 | 8.47488839285714 | 8.47488839285714 | 1.25498511904762 | 1.29035218253968 |
| 8 | 9.64145585317461 | 9.64145585317461 | 1.18425099206349 | 1.21961805555556 |
| 9 | 10.7372891865079 | 10.7372891865079 | 1.11351686507937 | 1.14888392857143 |
| 10 | 11.7623883928571 | 11.7623883928571 | 1.04278273809524 | 1.07814980158730 |
| 11 | 12.7167534722222 | 12.7167534722222 | 0.972048611111111 | 1.00741567460317 |
| 12 | 13.6003844246032 | 13.6003844246032 | 0.901314484126984 | 0.936681547619048 |
| 13 | 14.4132812500000 | 14.4132812500000 | 0.830580357142857 | 0.865947420634921 |
| 14 | 15.1554439484127 | 15.1554439484127 | 0.759846230158730 | 0.795213293650794 |
| 15 LSB | 15.5000000000000 | 15.5000000000000 | 0.689112103174603 | 0.724479166666667 |

**Extended Data Figure 5**: Illustration of Intra-path delay redistribution (I-PDR) technique: (a) an exemplary path of logic gates, (b) a bar plot of corresponding final redistributed gate delays, where all gates had equal delay $T$ before the redistribution.

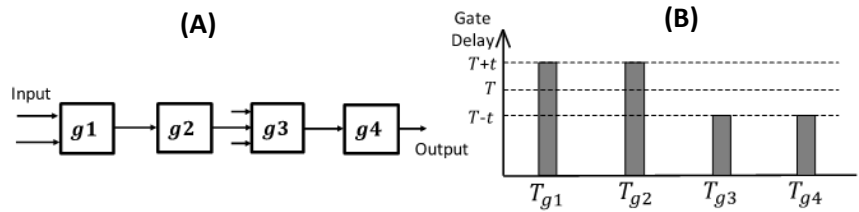

# Supplementary Information:

# Error Statistics Shaping via I-PDB and I-PDR techniques

In this section, we describe general algorithms for I-PDB and I-PDR, and explain how those can be applied to achieve error statistics shaping for any given logic network.

**Preliminary Definitions and Notation:**

Consider a combinational logic network ($S$) having a $l$-bit output, consisting of ASL gates. Let total number of gates be $N_g$ and each gate be denoted as $g_j, j \in \{1, \ldots N_g\}$. Let the delay of each gate be denoted as $T_{g_j}$ and its switching energy consumption as $E_{g_j}$. For the given logic network $S$, the total switching energy consumption $\mathcal{E}$ to compute one $l$-bit output word is given as:

$$\mathcal{E}\left(K_{g_1}, \ldots, K_{g_{N_g}}, T_{g_1}, \ldots, T_{g_{N_g}}\right) = \sum_{j=1}^{N_g} E_{g_j} = \sum_{j=1}^{N_g} \frac{K_{g_j}}{T_{g_j}}$$

Since energy can always be reduced at the expense of higher delay, we enforce an additional throughput constraint ($T_{th}$) at the system-level. Hence all output bits need to be computed within $T_{th}$ delay after the input bits are available.

In combinatorial circuits, there is no feedback. When each gate is considered as a node and each connection as an edge, the combinatorial circuit forms a directed acyclic graph (DAG). Thus $S = (V, E)$, where $V$ is a set nodes (gates) and $E$ is set of edges (connections). While each logic gate has its own input and output, we refer to the input and outputs of the network as primary inputs and primary outputs and assume that they are latched at a single clock edge. All the logic gates having one of the primary input as their input are referred to as primary input nodes. Similarly, primary output nodes are the logic gates whose output is the primary output.

We define a primary path in this DAG as a chain of cascaded nodes starting at one of the primary input nodes and terminating at one of the primary output nodes, while a path as any chain of cascaded nodes. Each path has a source node (a logic gate at the starting of the path) and a destination node (the last logic gate in the path). Each primary path is a subgraph $\rho_n = (G_n, C_n)$ and hence has an associated set of nodes $G_n \subset V$ defined as follows:

$$G_n = \{g_i | g_i \text{ is in path } \rho_n \: \forall \: i \} \: \forall \: n \in \{1, \ldots, N_p\}.$$

where $N_p$ is total number of primary paths. Similarly, $C_n$ is a set of connections between gates $g_i$ that constitute the primary path $\rho_n$. Let $m = \arg\max_n |G_n|$. Then $\rho_m$ is referred to as a critical path. Without loss of generality, let us assume that $\rho_1$ is a critical path, i.e. $\rho_1 = \rho_{cp}$. We define path delay as sum of the delays of the nodes in the path. We denote critical path delay by $T_{cp}$. If all nodes have equal delays, $T_{cp}$ is the maximum path delay.

We define all set operations on paths as corresponding operations on their individual node and edge sets. For example, $\rho_n \cap \rho_m \triangleq (G_n \cap G_m, C_n \cap C_m)$, $\rho_n \subset \rho_m \triangleq (G_n \subset G_m, C_n \subset C_m)$

Each primary path $\rho_n$ can be partitioned into $L$ disjoint subsets as follows:

$$\rho_n = \{\rho_{n,1} | \rho_{n,2} | \ldots | \rho_{n,L}\}$$

where each partition $\rho_{n,l} \: \forall \: l \in \{1, \ldots, L\}$ has following properties

(1) Each partition $\rho_{n,l}$ is a path that entirely lies in the primary path $\rho_n$.

(2) Either $\rho_{n,l} \subset \rho_1$ or $\rho_{n,l} \cap \rho_1 = \phi$ (empty set)

Without loss of generality, let us assume that partitions are indexed such that output of path $\rho_{n,1}$ is primary output and output of each path $\rho_{n,l}$ is input to path $\rho_{n,l-1}$. We refer to this indexing as 'ordered indexing'.

***Lemma 1***: The number of partitions of a primary path $\rho_n$ is minimized to $L$ if the following condition is satisfied:

If $\rho_{n,l} \subset \rho_1$ then $\rho_{n,l+1} \cap \rho_1 = \phi \: \forall \: l \in \{1, \ldots, L-1\}$

***Proof***:

By contradiction, suppose $\rho_{n,l} \subset \rho_1$ and $\rho_{n,l+1} \cap \rho_1 \neq \phi$ for some $l$.

By the above property (2), we claim that $\rho_{n,l+1} \subset \rho_1$.

Then, minimum number of partitions will be reduced to $L-1$ since new partition $\rho_{n,l}^* = (\rho_{n,l} \cup \rho_{n,l+1})$ can be defined by merging $\rho_{n,l}$ and $\rho_{n,l+1}$.

Now we define an input critical path ($\rho_S^{\text{imax}}(g_i)$) for each node $g_i$ in DAG $S$ as the path having the largest number of nodes, such that its source node is a primary input node and its destination node is the node $g_i$. Similarly, output critical path ($\rho_S^{\text{omax}}(g_i)$) for node $g_i$ can be defined as the path having the highest number of nodes, such that its source node is $g_i$ and destination node is a primary output node. If there are multiple input (output) critical paths for a certain gate $g_i$, they are denoted as $\rho_{j,S}^{\text{imax}}(g_i)$ $\left(\rho_{j,S}^{\text{omax}}(g_i)\right)$ with index $j$ used to distinguish between them. These terminologies will be used in subsequent subsections to describe I-PDB and I-PDR algorithms that can be applied to any given logic network as well as their properties.

**I-PDB algorithm and its properties:**

In I-PDB, the delays of gates on the non-critical path are increased such that every gate lies on at least one path having delay equal to the critical path delay, i.e. $T_{cp}$. After applying I-PDB, we refer to the resulting delay assignment as a balanced delay assignment. Its mathematical definition is as follows:

**Definition 1:** Any given delay assignment $\left(T'_{g_1}, \ldots, T'_{g_{N_g}}\right)$ is referred to as a *balanced delay assignment* if, for every gate $g_i$, its gate delay $T'_{g_i}$ is given as:

$$T'_{g_i} = T_{cp} - T_S^{\text{imax}}(g_i) - T_S^{\text{omax}}(g_i) + 2T'_{g_i}$$
$$T'_{g_i} = T_S^{\text{imax}}(g_i) + T_S^{\text{omax}}(g_i) - T_{cp} \tag{A7}$$

where $T_S^{\text{imax}}(g_i)$, and $T_S^{\text{omax}}(g_i)$ are the delay of $\rho_S^{\text{imax}}(g_i)$ and $\rho_S^{\text{omax}}(g_i)$ respectively.

If a delay assignment of a logic network is balanced, it is impossible to increase the delay of a gate without decreasing the delay of other gate. There are many possible balanced delay assignments for a given logic network.

For the logic networks consisting of large number of gates, finding balanced delay state directly by inspection may not be possible. Hence, we derive a general I-PDB algorithm that can be applied to any logic network. This algorithm finds a balanced delay assignment starting with a delay assignment having all gate delays equal and updating individual gate delays. Now Lemma 2 investigates the properties of

$\rho_S^{\text{imax}}(g_i)$ and $\rho_S^{\text{omax}}(g_i)$ for any gate $g_i$ in order to determine one particular sequence for updating gate delays such that eventual gate delay updates do not violate the condition in ($A7$) for gates whose delays have been assigned/updated earlier. Thus, one can find balanced delay assignment in one iteration of updating individual gate delays in I-PDB algorithm.

***Lemma 2:*** Let the delays of all gates be equal and gate $g_i$ be any gate in a path $\rho_n$ in the logic network $S$. Then,

$$\rho_S^{\text{imax}}(g_i) \subset \varrho(\rho_n) \text{ and } \rho_S^{\text{omax}}(g_i) \subset \varrho(\rho_n)$$

where a set of paths $\varrho(\rho_n)$ is defined as

$$\varrho(\rho_n) = \left\{ \text{primary path } \rho_r \text{ such that } |G_r| \geq |G_n| \ \forall \ r \in \{1, \ldots, N_p\} \right\}.$$

***Proof:***

First, we will prove $\rho_S^{\text{imax}}(g_i) \subset \varrho(\rho_n)$:

$$\rho_S^{\text{imax}}(g_i) \subset \varrho(\rho_n) \equiv \left[ \exists k \text{ such that } \rho_k \in \varrho(\rho_n), G_S^{\text{imax}}(g_i) \subset G_k \text{ and } C_S^{\text{imax}}(g_i) \subset C_k \right]$$

Suppose that $\rho_S^{\text{imax}}(g_i) \not\subset \varrho(\rho_n)$. Let the primary path $\rho_m$ be a concatenation of path $\rho_S^{\text{imax}}(g_i)$ and path $\{\rho_S^{\text{omax}}(g_i) \backslash g_i\}$. Thus, $\rho_m \notin \varrho(\rho_n)$.

Now $|G_m|$ is the maximum value among all the paths that contain gate $g_i$ by definitions of input critical path and output critical path. Hence, $|G_m| \geq |G_n|$.

Thus, we have, $\rho_m \notin \varrho(\rho_n)$ even though $|G_m| \geq |G_n|$. This contradicts the definition of $\varrho(\rho_n)$. Hence, we prove that $\rho_S^{\text{imax}}(g_i) \subset \varrho(\rho_n)$.

If there are multiple input or output critical paths for the gate $g_i$, lemma 2 holds for each of them. Without loss of generality, let us index primary paths such that $|G_1| \geq |G_2| \geq |G_3| \geq \cdots \geq |G_{N_p}|$. If, for any $m$ and $n$, $|G_m| = |G_n|$, then $n < m$ if $|G_1 \cap G_n| > |G_1 \cap G_m|$

Let us assume that there are total $N_{cp}$ critical paths in a given logic network $S$. Now, PDB algorithm is applied to all the gates in the non-critical path as follows:

**Algorithm 1:**

**Input**: logic network $S$

**Output**: Balanced delay assignment $T_{g_i}\ \forall\ i$

$T_{g_i} \leftarrow 1\ \forall\ i$

$\varrho_{cp} = \bigcup_{r=1}^{N_{cp}} \rho_r$

$T_{cp} = |G_1|$

**for** $n = N_{cp} + 1:1:N_p$

    Partition $\rho_n = \{\rho_{n,1}|\rho_{n,2}|\ldots|\rho_{n,L}\}$

    $G_{cu} = \bigcup_{r=1}^{n-1} G_r$

    **for** $k = L:-1:1$

        **if** $(\rho_{n,k} \cap \varrho_{cp}) = \phi$

            **for** $q = 1:1:|G_{n,k}|$

                **if** $g_q \in (G_{n,k} \cap G_{cu}^c)$,

                    compute $T_{\varrho(\rho_n)}^{\text{imax}}(g_q), T_{\varrho(\rho_n)}^{\text{omax}}(g_q)$

                    $T_{g_q} = T_{\varrho(\rho_n)}^{\text{imax}}(g_q) + T_{\varrho(\rho_n)}^{\text{omax}}(g_q) - T_{cp}$

                **end**

            **end**

        **end**

    **end**

**end**

Balanced delay state puts the condition ($A7$) on the delay of every gate and delays of its input and output critical paths. When individual primary paths are considered in sequence, Lemma 2 establishes that input and output critical paths for any gate on a given path $\rho_n$ lie on one of the paths having more or equal number of gates than the path $\rho_n$ itself. This fact and condition ($A7$) are used together to derive delay update rule in the above I-PDB algorithm.

**Remark 1**: Suppose the delay of a given gate is increased by a factor of $\chi$ via I-PDB, the switching energy of that gate can be kept constant by reducing its supply current by a factor of $\sqrt{\chi}$. Even then, the energy delay product $K_g$ increases by the factor of $\chi$ reducing the switching error rate $\epsilon(K_g)$ of that gate as indicated in equation $(A6)$.

**I-PDR algorithm and its properties:**

Now we show how I-PDR can be applied to find any other balanced delay state starting with one balanced delay state.

In I-PDR, the gate delays along any path are redistributed such that the total path delay remains same. For example, as illustrated in ED figure 5A, if all gates have equal delay $T$, corresponding path delay will be $4T$. However, the gate delays can be redistributed such that gates $g_1$ and $g_2$ are operated at delay $T+t$ while gates $g_3$ and $g_4$ at $T-t$ (ED figure 5B), thus maintaining the path delay equal to $4T$. The choice of the value of $t$, in this case, controls the extent of delay redistribution.

A general rule of applying I-PDR systematically to any given logic network can be stated as follows:

**Algorithm 2:**

**Input:** balanced day assignment $T_{g_i} \ \forall \ i$

**Output:** another balanced delay assignment $T^*_{g_i} \ \forall \ i$

- Select a path $\rho_k$ and a gate $g_n$ on that path
- $T^*_{g_n} \leftarrow T_{g_n} - T_1$.
- Find gates $g_{q_1}, \dots, g_{q_N}$ such that for each gate $g_{q_i}$ one of the following two conditions holds,

$$[g_{q_i} \in G^{imax}_{i,S}(g_n) \text{ and } g_n \in G^{omax}_{S}(g_{q_i})]$$

or

$$[g_{q_i} \in G^{omax}_{i,S}(g_n) \text{ and } g_n \in G^{imax}_{S}(g_{q_i})]$$

- For each gate $g_{q_i}$, assign $T^*_{g_{q_i}} \leftarrow T_{g_{q_i}} + T_1$.

Thus, above procedure allows one to redistribute delays while maintaining total path delay constant.

**Remark 2:** If all gates are operating at the same supply current, the application of I-PDR does not change the total energy consumption of the logic network. Analogously, if all the gates are operating at the same delay, similar redistribution can be made for the supply current i.e., some gates can be operated at lower supply current and others at higher supply current, while making sure that the total energy consumption remains same.

**Error statistics shaping:**

In this work, the I-PDR is selectively applied such that the error distribution at the output of the logic network becomes sparse. When I-PDB is applied, the gates on all non-critical paths are slowed down and hence are made to operate at lower error rate. Then, I-PDR can be applied to make outputs of certain paths more error prone than others by reducing delays of appropriate gates and distributing them among others. For example, in ED Fig. 5, the outputs of gates $g_1$ and $g_2$ are less error prone than the output of gate $g_4$. Once we have such control over the error rates of outputs of different paths, following insight can be used as an empirical guidance towards achieving sparse error PMF.

Let us denote the error prone output of the logic network by $y_a$. We define correct output as the output of the logic network when $\epsilon=0$ for all logic gates, i.e. when all the logic gates are reliable. The correct output is denoted by $y_o$. The error prone output $y_a$ can be represented, in 2's complement form, as

$$y_a = -y_{a,0} + \sum_{i=1}^{l-1} 2^{-i} y_{a,i} \qquad (A8)$$

where $y_{a,i} \in \{0,1\}$ denotes $i$th bit of $y_a$ and $y_{a,0}$ denotes its sign bit. Denoting $i$th bit of the correct output by $y_{o,i}$, one can modify equation ($A8$) as follows:

$$y_a = -(y_{o,0} \oplus \beta_0) + \sum_{i=1}^{l} 2^{-i}(y_{o,i} \oplus \beta_i) \qquad (A9)$$

where $\beta_i$ is a Bernoulli random variable $\forall i \in \{0,1,\ldots,l\}$ that captures the impact of $\epsilon$-noisy gates on the output bits. It is to be noted that $\Pr\{\beta_i = 1\} = \Pr\{i\text{th output bit is in error}\}$. Simplifying equation (A9), we get,

$$y_a = \left(-y_{o,0} + \sum_{i=1}^{l} 2^{-i} y_{o,i}\right) + \left(-\beta_0(-1)^{y_{o,0}} + \sum_{i=1}^{l} 2^{-i}\beta_i(-1)^{y_{o,i}}\right) \quad (A10)$$

The second term in the equation $(A10)$ is the additive error in the correct output due of the component gates being $\epsilon$-noisy.

In order to shape the additive error distribution to be sparse, we make high magnitude errors to be more probable than low magnitude errors. This is achieved by meeting the following condition:

$$\Pr\{\beta_i = 1\} \; \forall \; i \in \{0,1,2,p-1\} \gg \Pr\{\beta_j = 1\} \; \forall \; i \in \{k+1, \dots, l\}, \quad p < l \quad (A11)$$

where $p$ is the numbers of most significant bits (MSBs) for which the error rate is chosen to be high.

The condition in equation (A11) indicates that for any logic network having multi-bit output, the sparse output error PMF can always be achieved by choosing appropriate error rates for its output bits. We apply I-PDB and I-PDR such that the MSBs of the output become more error prone that the LSBs in order to make output error PMF sparse.